\documentclass[aps,a4paper,showpacs,superscriptaddress]{revtex4}
\usepackage{subfigure}
\usepackage{graphicx}
\usepackage[all]{xy}
\usepackage{amsmath}
\usepackage{amssymb}
\usepackage{enumerate}
\usepackage{epsf} 
\newcommand{\be}{\begin{equation}}
\newcommand{\ee}{\end{equation}}
\newcommand{\ben}{\begin{eqnarray}}
\newcommand{\een}{\end{eqnarray}}
\newcommand{\bes}{\begin{subequations}}
\newcommand{\ees}{\end{subequations}}

\newcommand{\bb}{\bibitem}

\newcommand{\sech}{{\rm sech}}

\begin{document}
\title{Stable static structures in models with higher-order derivatives}
\author{D. Bazeia\footnote{bazeia@fisica.ufpb.br (corresponding author)}}
\affiliation{Departamento de F\'\i sica, Universidade Federal da Para\'\i ba 58051-970 Jo\~ao Pessoa, Para\'\i ba, Brazil}
\affiliation{Departamento de F\'\i sica, Universidade Federal de Campina Grande, 58109-970 Campina Grande, PB, Brazil}
\author{ A.S. Lob\~ao Jr.}
\affiliation{Departamento de F\'\i sica, Universidade Federal da Para\'\i ba 58051-970 Jo\~ao Pessoa, Para\'\i ba, Brazil}
\author{R. Menezes}
\affiliation{Departamento de F\'\i sica, Universidade Federal de Campina Grande, 58109-970 Campina Grande, PB, Brazil}
\affiliation{Departamento de Ci\^encias Exatas, Universidade Federal da Para\'\i ba, 58297-000 Rio Tinto, PB, Brazil}

\begin{abstract}
We investigate the presence of static solutions in generalized models described by a real scalar field in four-dimensional space-time.
We study models in which the scalar field engenders higher-order derivatives and spontaneous symmetry breaking, inducing the presence of domain walls. Despite the presence of higher-order derivatives, the models keep to equations of motion second-order differential equations, so we focus on the presence of first-order equations that help us to obtain analytical solutions and investigate linear stability on general grounds. We then illustrate the general results with some specific examples, showing that the domain wall may become compact and that the zero mode may split. Moreover, if the model is further generalized to include k-field behavior, it may contribute to split the static structure itself.
\end{abstract}

\pacs{11.27.+d}
\maketitle

\section{introduction}

In this work we deal with relativistic models described by a single real scalar field with generalized dynamics in four-dimensional space-time. The study is inspired on the Galileon field, which is a real scalar field that engenders Galilean invariance, that is, if $\pi=\pi(x)$ is real, it is a Galileon field if its Lagrange density is symmetric under the Galilean and shift transformation $\pi\to\pi+a {\cdot} x+b$, with $a$ being a constant vector and $b$ a constant scalar. 

The Galileon field was studied in \cite{g1,g2} aimed to investigate self-accelerating solutions in the absence of ghosts, and has been further investigated in a diversity of contexts, with direct phenomenological applications, as one can see in the recent reviews \cite{r1,r2,r3}. In particular, in \cite{gs1,gs2,gs3,gs4,gsuper1,gsuper2} the authors deal with solitonic solutions and supersymmetrization. In \cite{gs1} it is shown that the Galileon field cannot give rise to static solitonic solutions; however, in \cite{gs2} one investigates the presence of soliton-like traveling waves for the Galileon field in two-dimensional space-time. Also, in Refs.~\cite{gs3,gs4} the authors offer other interesting results on solitons and Galileons. In \cite{gsuper1}, supersymmetry is implemented starting from ghost condensate theories \cite{gsuper3}; see also Refs.~\cite{s1,s2,gs5} for other studies on supersymmetry, generalized models and integrability.

One motivation to study the Galileon field is inspired on the fact that the Galilean invariance is capable of inducing an important feature to the Galileon field, which keeps its equation of motion a second-order differential equation. This and the presence of supersymmetry suggest that we search for a first-order framework, that is, for first-order differential equation that solve the equation of motion. We shall do this, extending the model, using the Galileon field to control the kinematics, but adding other terms, which break the Galilean symmetry and allow for the presence of spontaneous symmetry breaking, giving rise to localized static solutions. We call the scalar field, generalized Galileon field. We remark that the Galilean symmetry forbids the appearance of static solutions \cite{gs1}, so we are forced to generalize the model, to break the Galilean symmetry to study the appearance of nontrivial static structures. Another motivation comes from gravity: we know that minimal coupling of Galileons to gravity leads to equations of motion which have higher-order derivatives of the metric; however, this can be remedied with non-minimal couplings, at the expense of breaking the Galilean symmetry \cite{DD}.

Here we focus attention on the model
\be
{\cal L}= K(\pi,X) + F(\pi,X) \Box \pi,
\ee
in four-dimensional space-time. We consider that $K(\pi,X)$ and $F(\pi,X)$ are in principle arbitrary functions of $\pi$ and $X$, with $X$ being defined as
\be
X=\frac12 \partial_\mu \pi \partial^\mu \pi.
\ee
We are using $\Box \equiv g^{\alpha\beta}\partial_\alpha\partial_\beta$, the metric is diagonal $(+,-,-,-)$ and the scalar field, space and time coordinates, and coupling constants are all dimensionless. Like in \cite{Kobayashi:2010cm,Deffayet:2010qz}, we change the term $\partial_\mu \pi \partial^\mu \pi \Box \pi$ to the more general form $F(\pi,X)\Box\pi$.

The generalization that we consider may break the Galilean symmetry, but the equation of motion preserves the second-order structure. We are interested in solutions of these theories in the presence of spontaneous symmetry breaking, and we shall search in particular for planar domain walls and for its classical stability. 
As one knows, domain walls are non-perturbative classical solutions which find applications in many areas in physics, describing transitions between disconnected states of minimum energy \cite{V,MS}. The main issue here is to study domain walls in models of scalar fields with generalized dynamics of the Galileon type. We may also include k-field dynamics \cite{B}, as we have done before in Refs.~\cite{Bazeia:2013yza,  Bazeia:2008tj, Avelino:2010bu,Bazeia:2013euc,Bazeia:2010vb}. Here we focus on similar issues, with the scalar field now having generalized dynamics. The results show that the Galileon-like field may make the static solution compact, and may split the zero mode. Moreover, if we add generalized kinematics to the dynamical field, making the scalar field a generalized k-Galileon, the two contributions may contribute to split the static structure itself.

The investigation is organized as follows. In the next two sections we introduce the model and study linear stability on general grounds. We focus in particular on the first-order framework, where we search for first-order ordinary differential equations whose solutions also solve the equation of motion, which is second-order ordinary differential equation. In Sec.~\ref{sec3} we employ the method in order to investigate some distinct models explicitly, searching for static solutions and showing that they may engender interesting features. We end the work in Sec.~\ref{sec4}, where we include our comments and conclusions.

\section{The model}\label{sec1}

We consider the case of a single real scalar field in four-dimensional space-time with action
\be\label{action}
{\cal S}=\int d^4x\,\left( K(\pi,X) + F(\pi,X) \Box \pi\right).
\ee
Here $K(\pi,X)$ and $F(\pi,X)$ are in principle generic functions, and we get the equation of motion
\ben\label{eqofmotiong}
\partial_\mu \left(K_X \partial^\mu\pi\right) - K_\pi+\partial_\mu \left(F_X S^\mu\right)-
\partial_\mu F_\pi \partial^\mu \pi -2 F_\pi \Box \pi = 0\,,
\een
where
\ben
S^\mu \equiv \Box \pi \partial^\mu \pi- \partial_\nu \pi\partial^\mu \partial^\nu \pi\,.
\een
We see that for a generic field configuration, the above equation of motion is second-order partial differential equation. 

We can use the general formulation for the energy-momentum tensor to obtain \cite{Moeller:2002vx} 
\ben\label{emtensor}
T_{\mu\nu}=&&-(K+F\Box \pi) g_{\mu\nu}+K_X \partial_\mu \pi \partial_\nu \pi+F_X \Box \pi \partial_\mu \pi \partial_\nu \pi -\partial_\mu F \partial_\nu \pi + F \partial_\mu \partial_\nu \pi\,.
\een 

Since we are interested in investigating domain walls, we suppose that the scalar field is static, that is, $\pi=\pi(x)$, such that  
\be\label{boundary}
\pi^{\prime} (x \to \pm \infty) \to 0\,,
\ee
where prime stands for derivative with respect to the spatial coordinate $x$.
In this case, $S^\mu$  vanish and the equation of motion \eqref{eqofmotiong} reduces to
\ben\label{secondorde}
\left[(K_X+2XK_{XX})-2(F_\pi+ X F_{\pi X})\right]\pi^{\prime\prime}
-2X(K_{\pi X}-F_{\pi\pi}) +K_\pi=0\,.
\een
It can be integrated once to give 
\begin{equation}\label{fistorde}
K-2XK_X +2X F_\pi=C\,,
\end{equation}
where $ C $ is constant of integration. This equation only depends on the first derivative of the scalar field, so it is a first-order differential equation. We note that if we take the derivative of \eqref{fistorde} with respect to $x$, we get back to \eqref{secondorde}, so the solutions of \eqref{fistorde} also solve the equation of motion.

For the static field $\pi(x)$, the only non-trivial components of the energy-momentum tensor \eqref{emtensor} are
\bes
\ben
\label{rho}T_{00}&=&-K + F \pi^{\prime\prime}\,, \\
T_{11}&=&K-2XK_X +2X F_\pi\,,
\een
\ees	
We use the first-order equation \eqref{fistorde} to see that the stress component of $T_{\mu\nu}$ is constant, that is, $T_{11}=C$. 

The total energy of the field configuration $\pi(x)$ can be obtained as
\be\label{energy}
E=\int^{\infty}_{-\infty} dx\,\left(-K(\pi,X) + F(\pi,X) \pi^{\prime\prime}\right)\,.
\ee
This expression is important to elaborate on stability, following the Derrick/Hobart scaling argument \cite{DH}, which introduces a necessary condition for the stability of the solution. To do this, we follow \cite{bmm} and introduce $\pi_\lambda(x) = \pi(\lambda x)$. We use $\pi_\lambda(x)$ to define $E_\lambda$ in the form
\be
E_\lambda=\int^{\infty}_{-\infty} dx\,\left(-K(\pi_\lambda,X_\lambda) + F(\pi_\lambda,X_\lambda) \pi_\lambda^{\prime\prime}\right)\,.
\ee
It leads to
\be\label{enerderic}
E_\lambda=\int^{\infty}_{-\infty} \!\!\!\!dx\,\left(-\lambda^{\!-1} K(\pi,\lambda^2 X) \!+ \!\lambda F(\pi,\lambda^2X) \pi^{\prime\prime}\right)\,.
\ee

We see that $ E_\lambda|_{\lambda\to1}\to E$, and so we search for 
\be\label{condictions}
\frac{\partial E_\lambda}{\partial \lambda}\bigg|_{\lambda\to1}\!\!\!\!\to0\,.
 \ee

This condition allows that we write equation \eqref{enerderic} as
\be
\int^{\infty}_{-\infty} \!\!\!\!\!\!dx\left(K(\pi,X)\! -\! 2K_X X \!-\! (2F_X X\!+\!F(\pi,X)) \pi^{\prime\prime}\right)\!=\!0.\;\;\;
\ee
We integrate by parts the last term and consider \eqref{boundary} to get
\be
\int^{\infty}_{-\infty} \!\!\!dx\,(K - 2K_X X + 2F_\pi X)=\int^{\infty}_{-\infty}\!\!\! dx \,T_{11}=0\,.
\ee
Since $T_{11}$ is constant, we then have to set $T_{11}=0.$ This extends the result obtained in Ref. \cite{Bazeia:2007df} to the present situation. It is the stressless condition, and it is necessary condition for stability of the static solution.

\section{Linear Stability}
\label{sec2}

In this section we investigate linear stability of the static solution. For completeness, we start investigating the behavior of the general solution of the equation of motion \eqref{eqofmotiong}.
We introduce general fluctuations for the  scalar fields in the form: $\pi(\vec x, t) = \pi(\vec x) + \eta(\vec x, t)$, where $ \pi(\vec x)$ represents the statical solution. In this case, up to first-order in the fluctuations we have
\bes
\begin{equation}
X\rightarrow X + \partial_\nu \pi \partial^\nu \eta\,,
\end{equation}
with this we get the contributions for $S^\mu$ as
\ben
S^\mu \rightarrow S^\mu+ M^{\mu\nu\alpha\beta} (\partial_\alpha\partial_\beta \pi \partial_\nu \eta+ \partial_\alpha\partial_\beta\eta\partial_\nu  \pi)\,,
\een
\ees
where $M^{\mu\nu\alpha\beta}=g^{\mu\nu} g^{\alpha\beta} -g^{\mu\alpha}g^{\nu\beta}$. After some algebraic manipulations, we can write 
\be\label{Eqofperturb}
\partial_\beta \left(A^{\alpha\beta}\partial_\alpha\eta\right) = B \eta\,,
\ee
where 
\bes
\ben
\!\!\!\!\!\!\!\!A^{\alpha\beta}(\vec x)\!\!\!&=&\!\!\!g^{\alpha\beta}K_X+K_{XX}\partial^\alpha\pi \partial^\beta\pi-2g^{\alpha\beta} F_\pi+M^{\mu\nu\alpha\beta}\left[F_X  \partial_\mu \partial_\nu \pi+\partial_\mu(F_X\partial_\nu \pi)\right]-F_{\pi X}\partial^\alpha\pi\partial^\beta\pi+F_{XX}\partial^\alpha\pi S^\beta\,,\;\;\;\\
\!\!\!\!\!\!\!\!B(\vec x)\!\!\!&=&\!\!\!K_{\pi\pi}-\partial_\mu( K_{\pi X }\partial^\mu\pi)- \partial_\mu\left(S^\mu F_{\pi X}\right)+\left(\partial_\mu F_{\pi\pi}\right)\partial^\mu \pi+2F_{\pi\pi}\square \pi\,.
\een
\ees

Despite the complexity of the above equation, it can be simplified for the specific case of planar domain wall, where $\pi=\pi(x)$. Here we get
\ben\label{Epert}
A^{00}\ddot \eta+\left(A^{11}\eta^{\prime}\right)^{\prime}+A^{ij}\partial_i\partial_j\eta = B \eta\,,
\een
for $i,j\neq 1$, where:
\bes
\ben
\!\!A^{00}\!\!&=&\!\!K_X-2F_\pi-F_X\pi^{\prime\prime} -\left(F_X\pi^\prime\right)^{\prime}\,;\\
\!\!A^{11}\!\!&=&\!\!-(K_X+2XK_{XX})+2(F_\pi+XF_{\pi X})\,;\\
\!\!A^{ij}\!\!&=&\!\!-\delta^{ij}A^{00}\,;\\
\!\!B\!&=&\!\!K_{\pi\pi}+( K_{\pi X }\pi^{\prime})^{\prime}-( F_{\pi\pi})^{\prime}\pi^{\prime}-2F_{\pi\pi} \pi^{\prime\prime}.\,\,
\een
\ees
We can separate variables and write the perturbation as
\be
\eta(\vec x,t)=\sum_n\xi_n(t,y,z) \psi_n(x)\,,
\ee
where 
\be
\xi_n(t,y,z)=\cos(w_n t)\cos(k_y y)\cos(k_z z).
\ee
Thus, the above equation \eqref{Epert} can be written as
\be\label{ppp1}
-\left(|A^{11}|\psi_n^{\prime}\right)^{\prime} =B\psi_n + A^{00}M_n^2\psi_n\,,
\ee
where $M_n^2=w_n^2-k_y^2-k_z^2$. In order to ease the investigation, we consider the case with $k_y=k_z=0$. It is appropriate to introduce new variables, and we make the changes
\be\label{chances}
dz=\frac{dx}{a(x)};\,\,\,\,\,\,\,\,\,\,\, \psi_n(x)=\frac{u_n(z)}{\sqrt{A^{00}a(x)}},
\ee
where
\be\label{stab}
a^2(x)=\frac{|A^{11}|}{A^{00}}\,.
\ee
This allows that we obtain the Schrodinger-like equation
\be\label{schroeq}
-(u_n)_{zz}+U(z)u_n=w_n^2 u_n\,,
\ee
where
\ben\label{Uquant}
U(z)&=&\frac{\left(\sqrt{A^{00}a}\right)_{zz}}{\sqrt{A^{00}a}}-\frac{1}{A^{00}}\left(K_{\pi\pi}+\frac1{a}\left( K_{\pi X }\frac{\pi_z}{a}\right)_{\!z}\right)+\frac{1}{A^{00}\pi_z}\left(\left(\frac{\pi_z}{a}\right)^2F_{\pi\pi}\right)_{\!z }\,,
\een
is the stability potential we have to solve to get the corresponding eigenvalues and eigenstates. We see that if $F$ vanishes, we get back to the result obtained in \cite{Bazeia:2008tj}. This is the general result, and we see that linear stability requires the eigenvalues $w_n^2$ to be non-negative. This depends crucially on the potential $U(z)$, which has to be investigated for each one of the specific models that we explore in the next section.

\section{Examples}\label{sec3}

Let us now investigate some specific models, to illustrate how the above investigation works for particular cases.

\subsection{Generalized Galileons}

We start with the case $K=0$. The model describes the generalized Galileon field, and the equation of motion reduces to 
\be
 (F_\pi)^\prime \pi^\prime + 2F_\pi \pi^{\prime\prime}=0\,,
\ee
which leads to the first-order equation $F_\pi X =C$. If we consider stressless solutions, we have to take $C=0$, and so there are no nontrivial localized static solutions in this case, in agreement with the results of Ref.~\cite{gs1}. Here we note that the necessary condition that comes from the Derrick/Hobart scaling argument very much simplifies the investigation on stability.

\subsection{Generalized Galileons and symmetry breaking}

Let us now study generalizations with $F(\pi, X)$, but supposing that $K(\pi,X)$ represents standard model, that is,
\be
K(\pi,X)=X-V(\pi).
\ee
In this case the first-order Eq.~\eqref{fistorde} can be written as
\begin{equation}\label{fistordes}
\pi^{\prime2}\left(1-2F_\pi\right)=2V\,,
\end{equation}
where we used $C=0$.
Note that if the potential $V(\pi)$ vanishes, we get back to the trivial result: the model supports no nontrivial localized static solutions.

\begin{figure}[t] 
\includegraphics[scale=0.7]{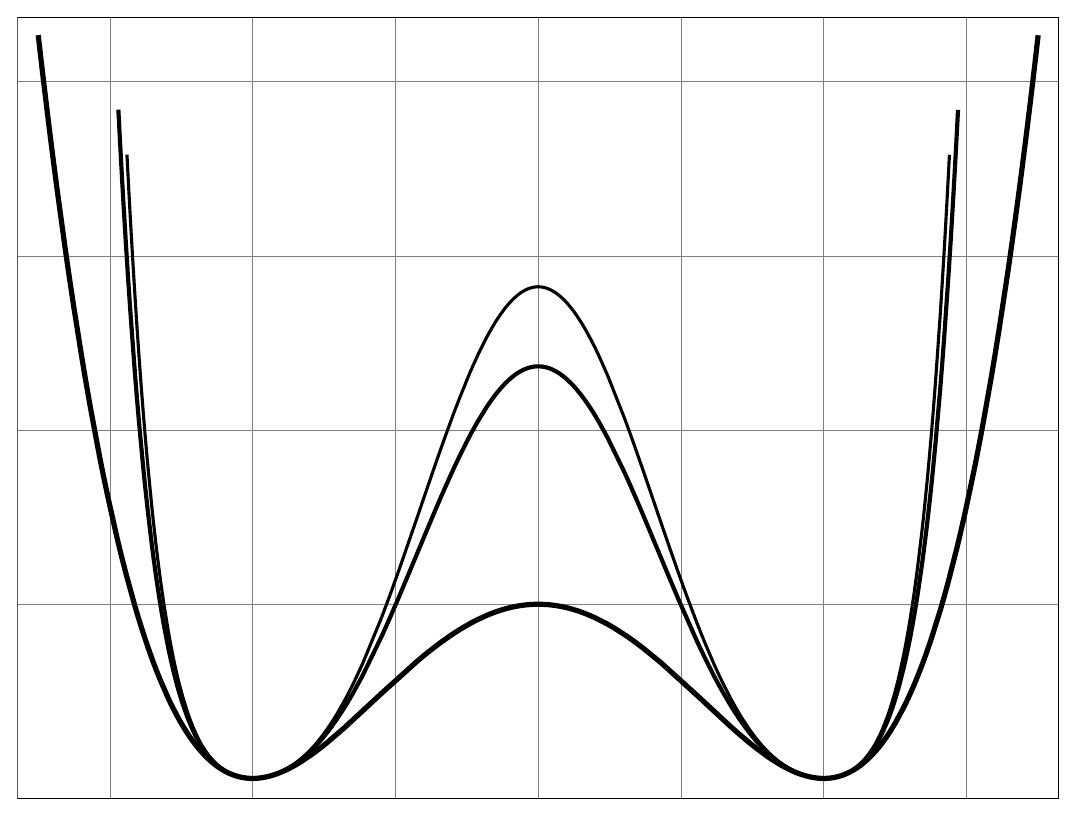}
\caption{\small{The potential \eqref{pot2}, plotted for $b=0$ (thicker line), $b=1.366$ (thick line), and $b=2$ (thinner line).}}
\label{figure1}
\end{figure}

\begin{figure}[t] 
\includegraphics[scale=0.7]{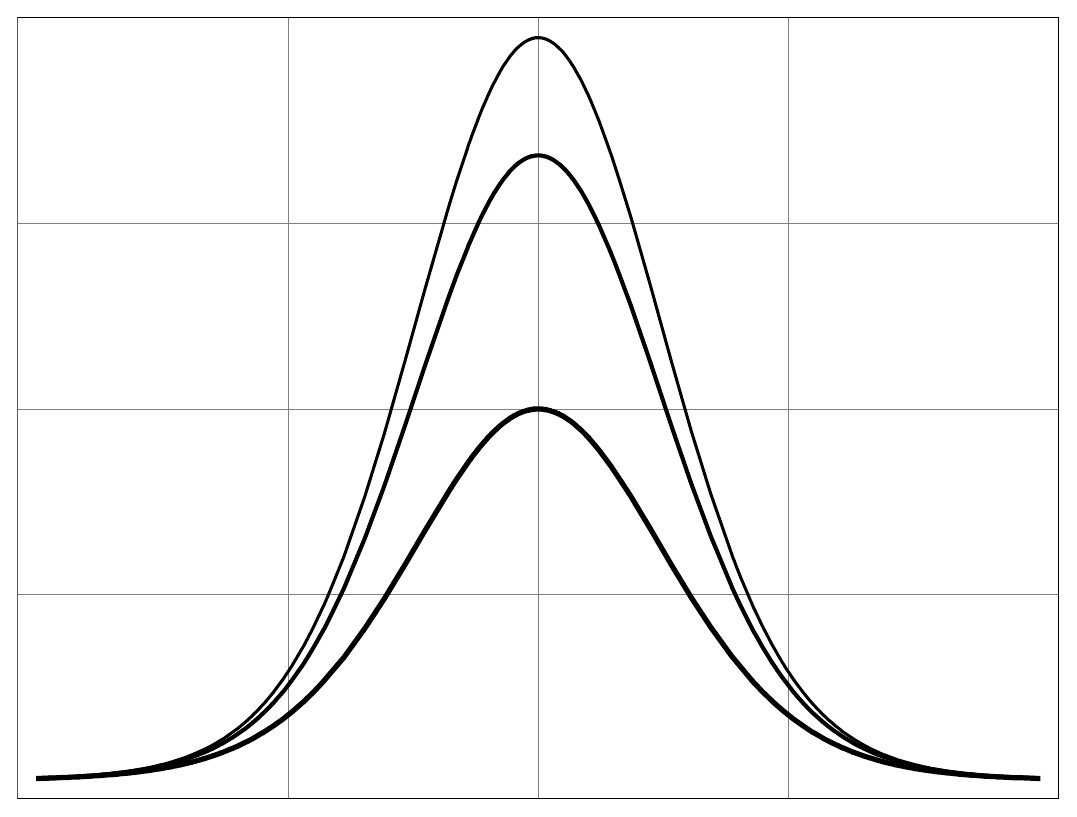}
\caption{\small{The energy density \eqref{enerdensity2}, plotted for $b=0$ (thicker line), $b=1.366$ (thick line), and $b=2$ (thinner line).}}
\label{figure2}
\end{figure}

If the potential does not vanish, we show explicitly that the model engenders nontrivial solutions if we take
\begin{equation}\label{model2}
F(\pi,X)=b X \pi\,.
\end{equation}
Here the equation \eqref{fistordes} becomes
\begin{equation}\label{fistorde2}
\pi^{\prime2}\left(1+b \pi^{\prime2}\right)=2V\,.
\end{equation}
It can be written as 
\be\label{foeq}
\pi^{\prime 2}=\frac{\sqrt{1+8bV(\pi)\,}-1}{2b},
\ee
which has to be investigated after specifying the potential. We note that the above equation is consistent with \eqref{fistorde2} in the limit $b\to0$. We also note that there is another solution of \eqref{fistorde2}, with the minus sign for the square root in \eqref{foeq}. It leads to imaginary configurations, an issue which is out of the scope of the current work. 
As an interesting example, we consider the potential in the form
\begin{equation}\label{pot2}
V(\pi)=\frac12\left(1-\pi^2\right)^2\left[1+b\left(1-\pi^2\right)^2\right]\,.
\end{equation}
It allows to solve the first-order equation analytically, with 
\be\label{sta}
\pi(x)=\tanh(x),
\ee
which is static solution we also have in the case $b=0$, in the standard model. Fig.~\ref{figure1} shows the behavior of the potential \eqref{pot2} for some values of $b$. The value $b=1.366$ is in fact $b=1/2+\sqrt{3}/2$, and is the value where the zero mode start to split; see Fig.~\ref{figure4}. In fact, there is another (negative) value of $b$, given by $1/2-\sqrt{3}/2$, where the zero mode also splits, but we will not consider it here. In the case of $b$ positive, the energy density becomes
\be\label{enerdensity2}
\rho(x)=S^4\left(1+b\,S^2-\frac12\, b\,S^4\right)\,.
\ee
where $S={\rm sech}(x)$. In Fig.~\ref{figure2} one shows the behavior of the energy density for some values of the parameter $b$. 

\begin{figure}[t] 
\includegraphics[scale=0.7]{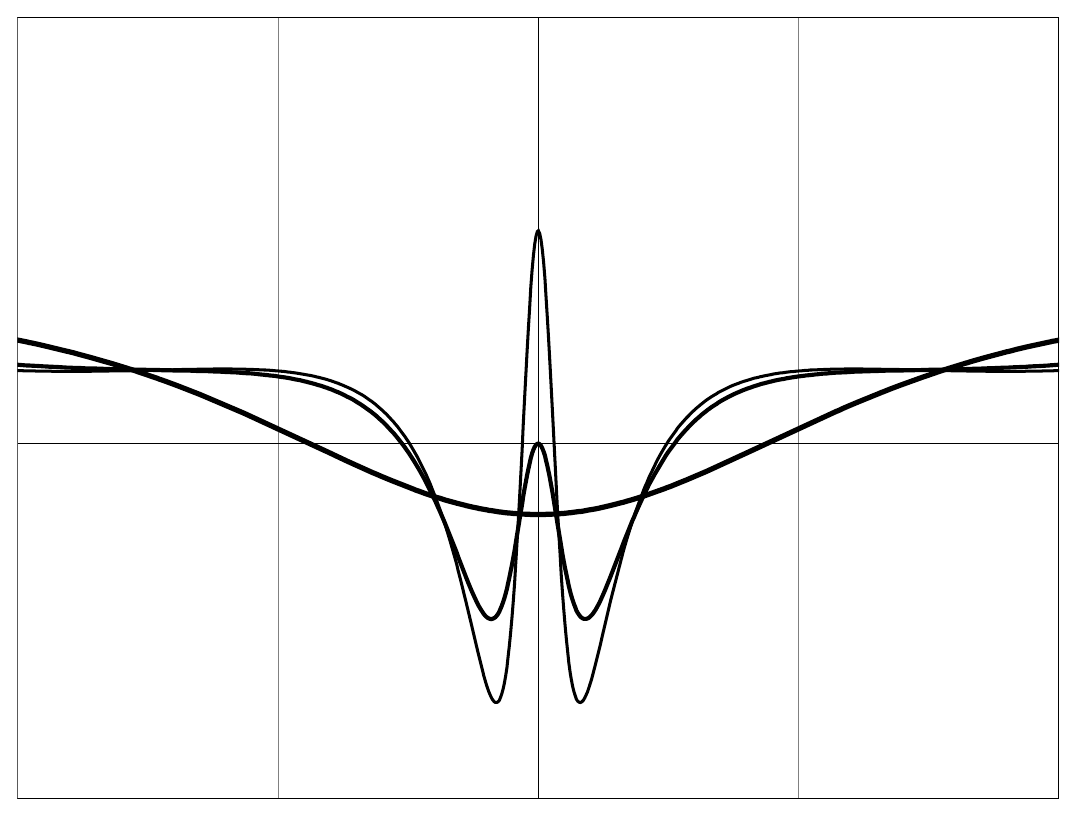}
\caption{\small{The stability potential, plotted for $b=0$ (thicker line), $b=1.366$ (thick line), and $b=2$ (thinner line).}}
\label{figure3}
\end{figure}

\begin{figure}[t] 
\includegraphics[scale=0.7]{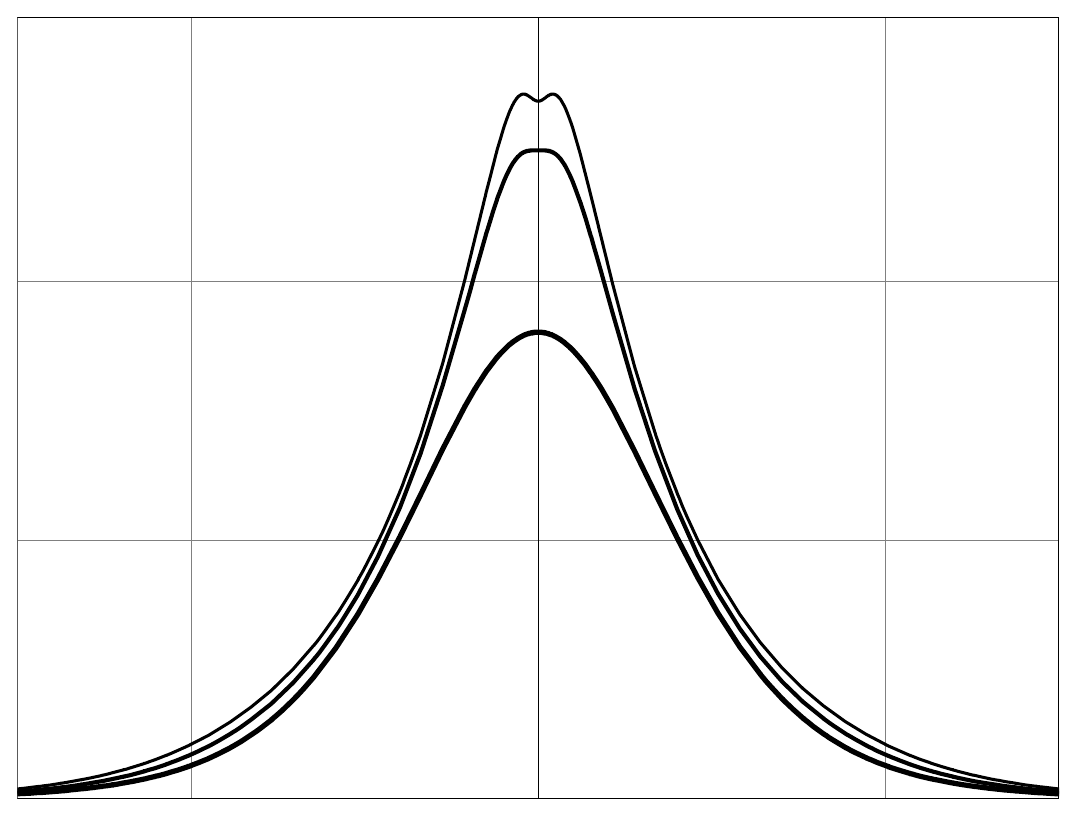}
\caption{\small{Plot of the zero mode for $b=0$ (thicker line), $b=1.366$ (thick line), and $b=2$ (thinner line).}}
\label{figure4}
\end{figure}

We go further and investigate linear stability of the model. We use Eq.~\eqref{ppp1} to get
\be
-\left[(1+2b\pi^{\prime2})\psi_n^{\prime}\right]^{\prime} =-V_{\pi\pi}\psi_n + \left(1-2b\pi\pi^{\prime\prime}\right)w_n^2\psi_n\,.
\ee
We can thus make the following change of variables
\bes
\be
dz\!=\sqrt{\!\frac{1+4 bS^2\left(1-S^2\right)}{1+2bS^4}\;}\;dx\,,
\ee
and
\be
u_n(z)\!=\![(1\!+\!4 bS^2(1\!-\! S^2))(1+2bS^4)]^{1/4}\psi_n\,.
\ee
\ees
In this case the parameter $b$ should be greater than $-1/2$.  The stability potential cannot be written analytically, but in Fig.~\ref{figure3} one shows how it behaves numerically for some values of $b$. It goes to the same value, $4$, as $z\to\pm\infty$, independently of $b$. We see that it engenders a double well behavior as $b$ increases, so we also investigate the zero mode, which is depicted in Fig.~\ref{figure4}. As expected, the zero mode responds to the double well behavior splitting, to accommodate itself into the two wells. This splitting of the zero mode is an interesting new behavior: it does not appear in the standard case, for $b=0$. To see how the splitting appears, we note from \eqref{schroeq} that the zero mode $u_0(z)$ obeys $-(u_0)_{zz}+U(z)u_0=0$. Moreover, in order to split, the zero mode has to change from a maximum to a local minimum at the origin, so it has to have an inflection point at $z=0$, such that $(u_0)_{zz}|_{z=0}=0$. As we see from the equation for the zero mode, this implies that the stability potential has to vanish at the origin, that is, $U(z=0)=0$. For the model under investigation, this is achieved for $b=1.366$, as we illustrate in Figs.~\ref{figure3} and \ref{figure4}.

Let us further study this model with another potential. We change \eqref{pot2} to the new form
\be\label{pot3}
V(\pi)=\frac12(1+b)(1-\pi^2)^{2}.
\ee
We use \eqref{foeq} to write
\be\label{eqb}
\pi^{\prime 2}=\frac{\sqrt{1+4b(1+b)(1-\pi^2)^{2}}-1}{2b},
\ee
which is consistent with \eqref{fistorde2} in the limit $b\to0$.


\begin{figure}[t] 
\includegraphics[scale=0.7]{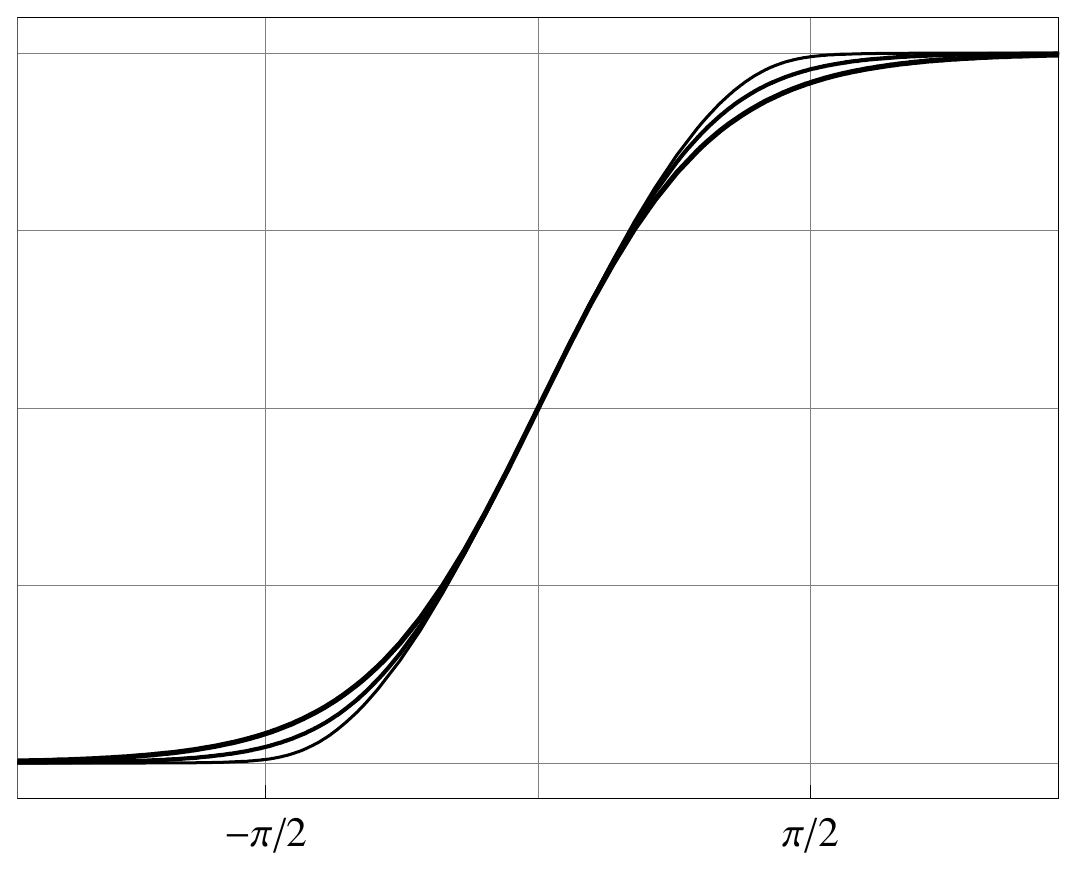}
\caption{\small{Plot of the static solution that solves \eqref{eqb} for $b=0$  (thicker line), $b=5$ (thick line), and $b=100$ (thinner line)}.}
\label{figure5}
\end{figure}

\begin{figure}[t] 
\includegraphics[scale=0.7]{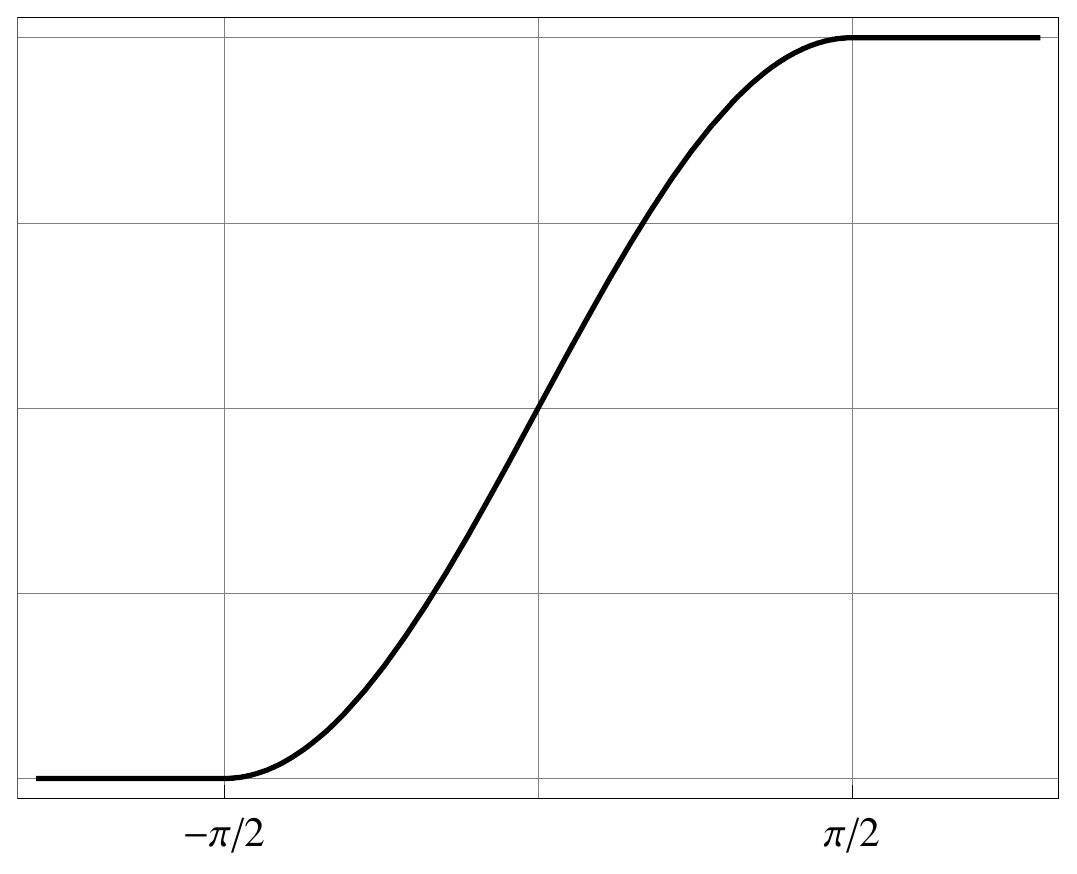}
\caption{\small{Plot of the static compact solution which appears for $b$ very large}.}
\label{figure6}
\end{figure}

We solve this equation numerically, and we plot the solution in Fig.~\ref{figure5}, for some values of $b$.
We note that as $b$ increases to very large values, the static solution shrinks, suggesting the appearance of a compact solution; see, e.g., Ref.~\cite{Bazeia:2010vb}. To see how the compact solution appears, we proceed as follows: we take $b$ very large, and from \eqref{eqb} we get
\be
\pi^\prime=|1-\pi^2|^{1/2}.
\ee
This equation is solved by
\begin{eqnarray}
\pi(x) = \left\{
\begin{array}{clc}
\sin(x) & \mbox{ for } &|x|\leq \pi/2\,, \\ 
 {\rm {sign}}(x) & \mbox{ for } & |x|> \pi/2\,.
\end{array} \right.
\end{eqnarray}
which is compact solution, which we depict in Fig.~\ref{figure6}. We see that the solution for $b=100$ in Fig.~\ref{figure5}
is essentially the compact solution plotted in Fig.~\ref{figure6}.

We also note that if one changes the potential \eqref{pot3} to the new form
\be
V(\pi)=\frac12(1+b)(1-\pi^2)^{4},
\ee
then in the limit of very large $b$ we get $\pi^\prime=(1-\pi^2)$.
This result leads us back to the standard solutions, described by Eq.~\eqref{sta}.

\subsection{Generalized k-Galileon and symmetry breaking}

Let us consider another model, now changing the $K(\pi,X)$ contribution to the generalized form
\begin{equation}\label{x2}
K=X+b X^2-V.
\end{equation}
This generalized form is sometimes called k-field; see, e.g., Ref.~\cite{Bazeia:2013yza,  Bazeia:2008tj, Avelino:2010bu,Bazeia:2013euc,Bazeia:2010vb}. This explains the term k-Galileon that we are using to name this subsection. Here we also take
\begin{equation}\label{f2}
F=\frac32b\pi X,
\end{equation}
which is essentially the same $F$ we have considered in the previous subsection; the factor $3/2$ is included in the above function for simplicity. We are adding the $X^2$ term in \eqref{x2} with the same parameter $b$, to simplify the first-order equation, as we show below.
We study no other possibility in this work.    

We use \eqref{x2} and \eqref{f2}, and now the first-order equation changes to 
\begin{equation}
\pi^{\prime 2}=2V.
\end{equation}
If we take the standard potential
\begin{equation}
V(\pi)=\frac12(1-\pi^2)^2,
\end{equation}
the solution becomes the standard one, $\pi(x)=\tanh(x)$, as in \eqref{sta}. Here, however, the energy density has the form
\begin{equation}
\rho(x)=S^4-\frac{b}{4}S^6 \Big(7S^2-6\Big).
\end{equation}
It has an inflection point at $x=0$, for $b=4/5$. Thus, for $b>4/5$ the energy density starts to split, indicating that the static structure engenders the interesting behavior of splitting itself. 

\begin{figure}[t]
\includegraphics[scale=0.7]{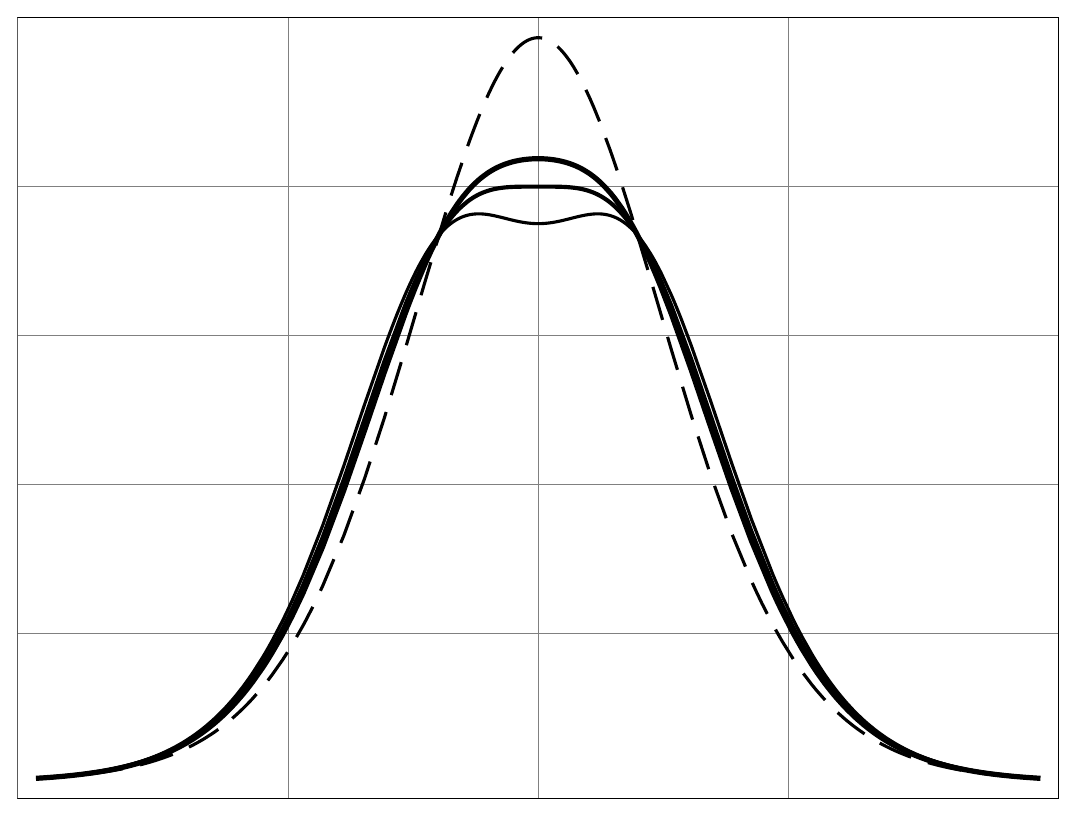}
\caption{{\small The energy density, depicted for $b=0$ (dashed line), $b=0.65$ (thicker line), $b=0.8$ (thick line), and $b=0.95$ (thinner line).}}
\label{figure7}
\end{figure}

\begin{figure}[ht]
\includegraphics[scale=0.7]{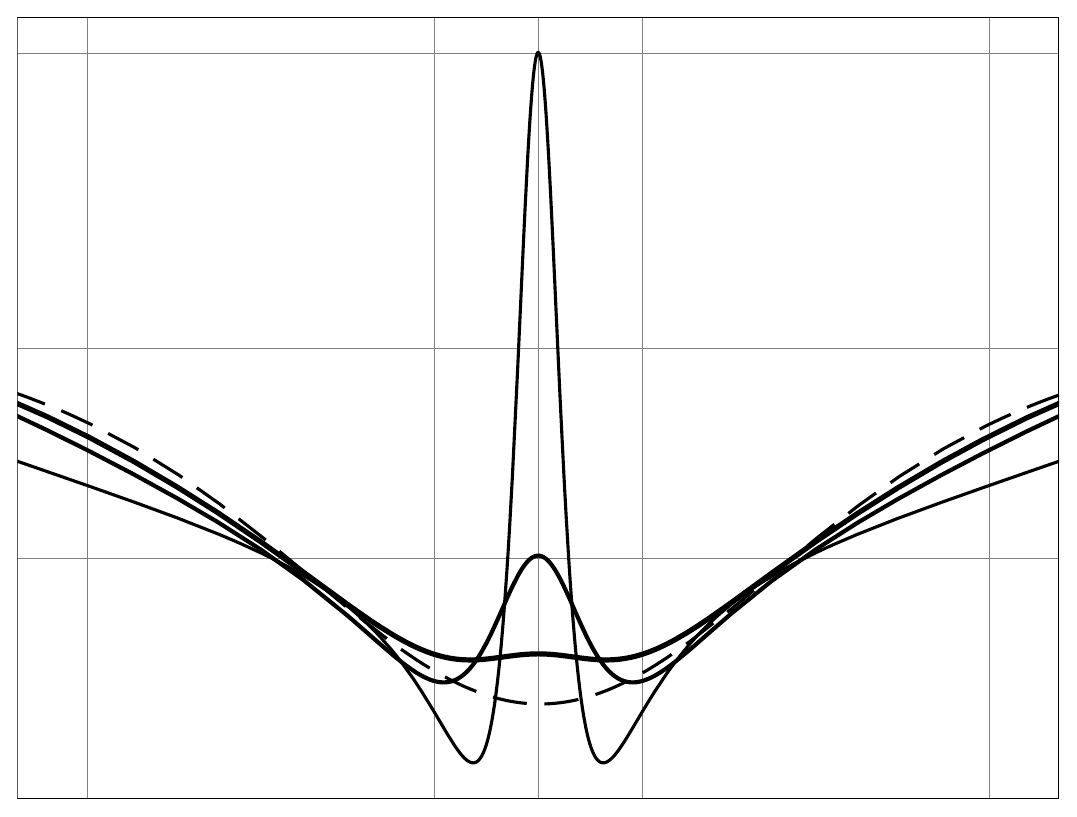}
\caption{{\small The stability potential, depicted for $b=0$ (dashed line), $b=1/6$ (thicker line), $b=1/3$ (thick line), and $b=2/3$ (thinner line).}}
\label{figure8}
\end{figure}

\begin{figure}[ht]
\includegraphics[scale=0.7]{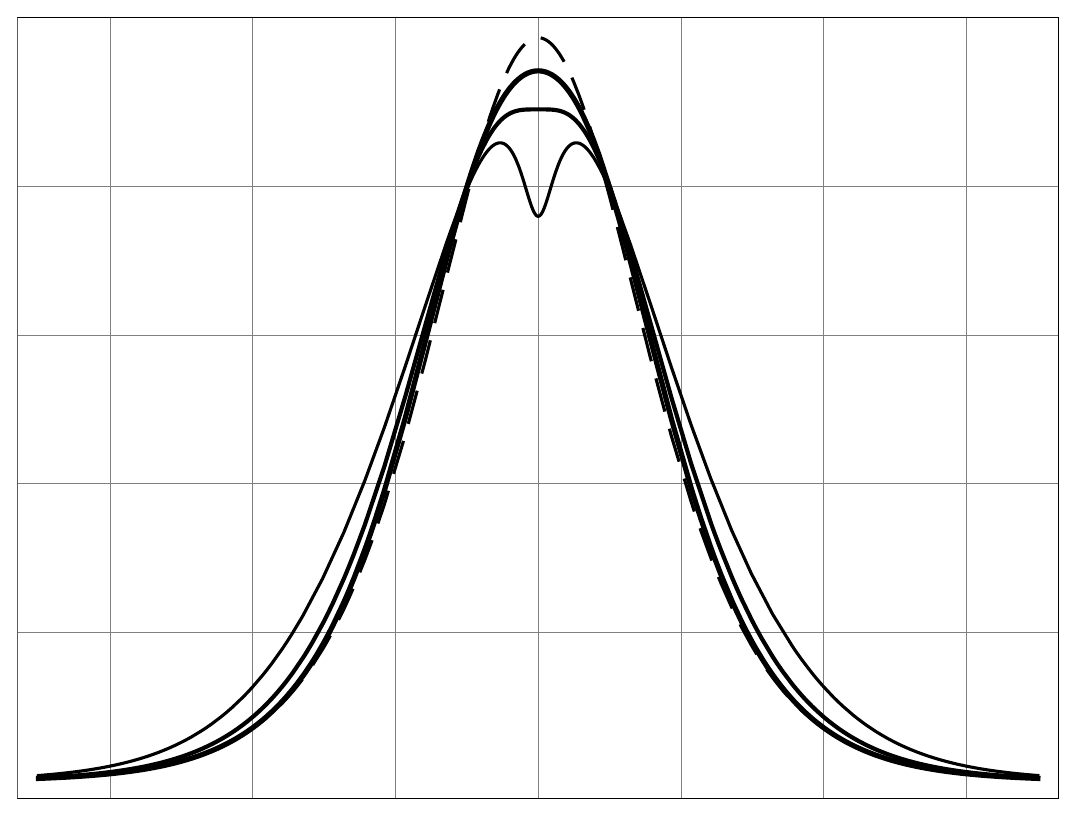}
\caption{{\small The zero mode, depicted for $b=0$ (dashed line), $b=1/6$ (thicker line), $b=1/3$ (thick line), and $b=2/3$ (thinner line).}}
\label{figure9}
\end{figure}

To see how the splitting appears, in Fig.~\ref{figure7} we plot the energy density for some values of $b$. Moreover, to study the behavior of the model under stability, we note that in this new model, stability leads us to

\begin{equation}
-\psi_n^{\prime\prime}=-V_{\pi\pi}\psi_n+\Big[1-b \left(\pi^{\prime 2}+3\pi\pi^{\prime\prime}\right)\Big]w_n^2\psi_n.
\end{equation}
To write the Schroedinger-like equation and identify the stability potential we take
\begin{equation}
dz=\sqrt{1-b\,S^2\Big(7S^2-6 \Big)}\,dx,
\end{equation} 
and
\begin{equation}
u_n(z)=\Big[1-b\,S^2\left(7 S^2-6 \right)\Big]^{1/4}\psi_n(x),
\end{equation}
which requires that $b<1$. In this case, the stability potential cannot be given explicitly, but we can depict it numerically, as we show in Fig.~\ref{figure8}. Also, we can investigate the zero mode numerically. The study shows that it also splits for $b\in(1/3,1)$, and in Fig.~\ref{figure9} we depict the zero mode for some values of $b$.

\begin{figure}[t]
\includegraphics[scale=0.7]{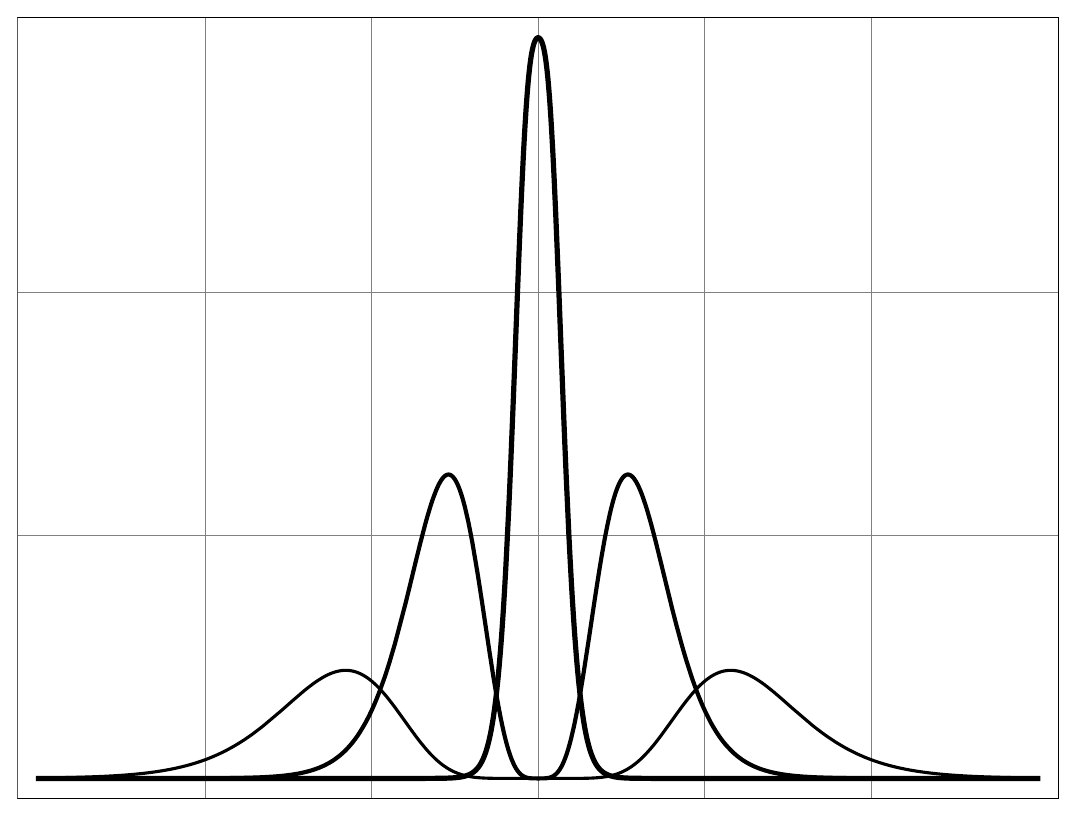}
\caption{{\small The energy density, depicted for $b=0.95$, and for $p=1$ (thinner line), $p=3$ (thick line), and $p=5$, thicker line.}}
\label{figure10}
\end{figure}

If we want to focus on the splitting of the static structure and make it more evident, we choose another, more appropriate model.
To illustrate this situation we follow \cite{S2}, and we introduce the potential
\be\label{pmodel}
V(\pi)=\frac12 \left(\pi^{\frac{p-1}{p}}-\pi^{\frac{p+1}{p}}\right)^2
\ee
where $p$ is an integer, odd, $p=1,3,5,...$. The case $p=1$ leads us back the previous model. For $p$ odd, arbitrary, the solution is
\be 
\pi(x)=\tanh^p(x/p); \;\;\;p=1,3,5,...,
\ee
and the energy density gets to the form
\ben
\rho(x)&=&S_p^4\;  T_p^{2 p-2}\Biggl(1+\frac{3b}{4}\;S_p^2\; T_p^{2 p}+b\;\Big(\frac{3-4p}{4p}\Big)\; S_p^4\;T_p^{2p-2}\;\Biggr),
\een
where $S_p=\sech(x/p)$ and $T_p=\tanh(x/p)$.
It depends on $b$ and $p$, and it is depicted in Fig.~\ref{figure10} for $b=0.95$ and for $p=1,3$ and $5$. The figure shows explicitly that the new paramer $p$ directly contributes to expand the splitting of the defect structure. Similar effects appear in the corresponding zero modes; the calculations follow the previous model, so we omit them here. 

The model \eqref{pmodel} is of interest, since one knows that the parameter $p$ mimics the presence of temperature, as it was introduced in another model \cite{S1}, described by a complex scalar field, used to split the brane in the braneworld scenario with a single extra dimension of infinite extent. See, e.g., Refs.~{\cite{S2,S1}} for further investigations on this issue.

\section{Conclusions}\label{sec4}

In this work we investigated the presence of localized static domain wall solutions in generalized models, described by the Galileon field, but enlarged
to accommodate spontaneous symmetry breaking to support localized static solutions. The study is implemented under the first-order framework, with the help of the Derrick/Hobart scaling argument and the stressless condition for stability.

The general investigation is then illustrated with some distinct models, from which we could construct stable domain wall configurations, having the form of the standard domain wall, which appears analytically as the hyperbolic tangent. Moreover, we could find compact solutions, depending on the way the scalar field self-interacts. In particular, we identified an interesting behavior: the splitting of the zero mode, controlled by $b$, the parameter that induces deviation of the model from the standard model, making the scalar field a Galileon-like field. The splitting of the zero mode may modify the scattering of static structures, and may contribute to change their collective behavior, a subject of current interest in high energy physics; see, e.g., \cite{Sc1,Sc2,Sc3} and references therein. 

We have investigated another model, in which one includes k-field kinematics and the Galileon-like behavior. We studied the case where the two effects cancel each other from the first-order equation, leaving it as in the standard model. However, they change the energy density and stability, and split the zero mode and the static solution itself. The splitting of the static structure is another interesting feature, which we think is generic and will remain in the braneworld scenario with a single extra dimension of infinite extent; see, e.g.,
Refs.~\cite{S2,S1,S3,S4,S5,S6}. This fact motivates us to investigate the models studied in this work minimally coupled to gravity, in the thick braneworld scenario with a single extra spatial dimension of infinite extent. We shall further report on this elsewhere.

The authors would like to thank CAPES and CNPq, for partial financial support.



\begin{thebibliography}{99}


\bb{g1}A. Nicolis, R. Rattazzi, and E. Trincherini, Phys. Rev.  D {\bf 79}, 064036 (2009).

\bb{g2}C. Deffayet, G. Esposito-Farese, and A. Vikman, Phys. Rev.  D {\bf79}, 084003 (2009).

\bb{r1}M. Trodden and K. Hinterbicher, Class. Quan. Grav. {\bf28}, 204003 (2011).

\bb{r2}C. de Ram, Comptes Rendus Physique {\bf 13}, 666 (2012).

\bb{r3}C. Deffayet, D.A. Steer, Class. Quant. Grav. {\bf30}, 214006 (2013).

\bb{gs1}S. Endlich, K. Hinterbichler, L. Hui, A. Nicolis, and J. Wang, JHEP {\bf1105}, 073 (2011).

\bb{gs2}A. Masoumi and Xiao Xiao, Phys. Lett. B {\bf715}, 214 (2012).
\bb{gs3}S.-Y. Zhou, Phys. Rev. D {\bf85}, 104005 (2012).
\bb{gs4}A. Padilla, P.M. Saffin, S.-Y. Zhou, Phys. Rev. D {\bf83}, 045009 (2011).


\bb{gsuper1}J. Khoury, J.L. Lehners, and B.A. Ovrut, Phys. Rev. D {\bf84}, 043521 (2011).
\bb{gsuper2}M. Koehn, J.L. Lehners, and B.A. Ovrut, {\it A Cosmological Super-Bounce}, arXiv:1302.0840.
\bb{gsuper3}J. Khoury, J.L. Lehners, and B. Ovrut, Phys. Rev. D {\bf83}, 125031 (2011).


\bb{s1}D. Bazeia, R. Menezes, A. Yu. Petrov, Phys. Lett. B {\bf683}, 335 (2010). 
\bb{s2}C. Adam, J.M. Queiruga, J. Sanchez-Guillen, and and A. Wereszczynski, Phys. Rev. D {\bf84}, 065032 (2011).
\bb{gs5}D. Bazeia, L. Losano, and J.R.L. Santos, {\it Solitonic traveling waves in Galileon theory.} arXiv:1408.3822.

\bb{DD}C. Deffayet, S. Deser, and G. Esposito-Farese, Phys. Rev. D {\bf80}, 064015 (2009).
\bibitem{Kobayashi:2010cm}
  T.~Kobayashi, M.~Yamaguchi and J.~Yokoyama, Phys. Rev. Lett. {\bf105}, 231302 (2010).

\bibitem{Deffayet:2010qz}
  C.~Deffayet, O.~Pujolas, I.~Sawicki and A.~Vikman, JCAP {\bf1010}, 026 (2010).

\bb{V}A. Vilenkin and E.P.S. Shellard, {\it Cosmic Strings and other topological defects} (Cambridge UP, Cambridge, UK, 1994).
\bb{MS}N. Manton and P. Sutcliffe, {\it Topological Solitons} (Cambridge UP, Cambridge, UK, 2004).

\bb{B}E. Babichev, Phys. Rev. D {\bf74}, 085004 (2006).

\bibitem{Bazeia:2008tj} 
  D.~Bazeia, L.~Losano and R.~Menezes, Phys.\ Lett.\ B {\bf 668}, 246 (2008).
 
\bibitem{Avelino:2010bu} 
  P.P.~Avelino, D.~Bazeia, R.~Menezes, and J.G.G.S. Ramos,  Eur.\ Phys.\ J.\ C {\bf 71}, 1683 (2011).

\bibitem{Bazeia:2013euc} 
  D.~Bazeia, A.~S.~Lob\~ao, L.~Losano and R.~Menezes,  Phys.\ Rev.\ D {\bf 88}, 045001 (2013).
  
\bibitem{Bazeia:2013yza}
  D.~Bazeia, A.~S.~Lob\~ao, L.~Losano and R.~Menezes, Eur. Phys. J. C {\bf74}, 2755 (2014).

\bibitem{Bazeia:2010vb} 
  D.~Bazeia, E.~da Hora, R.~Menezes, H.~P.~de Oliveira and C.~dos Santos, Phys.\ Rev.\ D {\bf 81}, 125016 (2010).
 
\bibitem{Moeller:2002vx}
  N.~Moeller and B.~Zwiebach,  JHEP {\bf 0210}, 034 (2002).

\bb{DH}R. Hobart, Proc. Phys. Soc. Lond. {\bf82}, 201 (1963); G.H. Derrick, J. Math. Phys. {\bf5}, 1252 (1964).

\bb{bmm}D.Bazeia, J. Menezes and R. Menezes, Phys. Rev. Lett. {\bf91}, 241601 (2003).
\bibitem{Bazeia:2007df}D.~Bazeia, L.~Losano, R.~Menezes and J.~C.~R.~Oliveira,  Eur.\ Phys.\ J.\  C {\bf 51}, 953 (2007). 
\bb{S2}D. Bazeia, C. Furtado, A.R. Gomes, JCAP {\bf0402}, 002 (2004).
\bb{S1}A. Campos, Phys. Rev. Lett. {\bf88}, 141602 (2002).
\bb{Sc1}P. Dorey, K. Mersh, T. Romanczukiewicz, and Y. Shnir, Phys. Rev. Lett. {\bf107}, 091602 (2011).
\bb{Sc2}M.A. Amin, E.A. Lim, and I-S. Yang, Phys. Rev. Lett. {\bf111}, 224101 (2013).
\bb{Sc3} A.R. Gomes, R. Menezes, K.Z. Nobrega and F.C. Simas, Phys. Rev. D {\bf90}, 065022 (2014).
\bb{S3}R.A.C. Correa, A. de Souza Dutra and M.B. Hott, Class. Quant. Grav. {\bf28}, 155012 (2011).
\bb{S4}Jie Yang, Yun-Liang Li, Yuan Zhong, Yang Li, Phys. Rev. D {\bf85}, 084033 (2012).
\bb{S5}D. Bazeia, A.S. Lob\~ao Jr., R. Menezes, A.Yu. Petrov, and A.J. da Silva, Phys. Lett. B {\bf729}, 127 (2014).
\bb{S6}W.T. Cruz, L.J.S. Sousa, R.V. Maluf, C.A.S. Almeida, Phys. Lett. B {\bf730}, 314 (2014).

\end{thebibliography}
 \end{document}